\documentstyle[12pt,epsfig]{article}
\begin{document}
\title{Detecting barriers to transport: \\
A review of different techniques}

\author{G.~Boffetta$^1$, G.~Lacorata$^{2,3}$, G.~Redaelli$^3$ 
and A.~Vulpiani$^4$}

\maketitle
\centerline{ $^1$Dipartimento di Fisica Generale and INFM, 
Universit\`a di Torino}
%\centerline{and INFM, Unit\`a di Torino Universit\`a,}
\centerline{ Via Pietro Giuria 1, 10125 Torino, Italy}
\centerline{and Istituto di Cosmogeofisica del CNR, 
Corso Fiume 4, 10133 Torino, Italy.}
\centerline{$^2$Dipartimento di Fisica, Universit\`a ``La Sapienza'',} 
\centerline{Piazzale Aldo Moro 5, 00185 Roma, Italy.}
\centerline{$^3$Dipartimento di Fisica, Universit\`a dell' Aquila}
\centerline{Via Vetoio 1, 67010 Coppito, L'Aquila, Italy.}
\centerline{$^4$Dipartimento di Fisica, Universit\`a ``La Sapienza''}
\centerline{and INFM, Unit\`a di Roma 1,}
\centerline{Piazzale Aldo Moro 5, 00185 Roma, Italy.}

\begin{abstract}
We review and discuss some different techniques for 
describing local dispersion properties in fluids. 
A recent Lagrangian diagnostics, based on the Finite Scale
Lyapunov Exponent (FSLE), is presented and compared to 
the Finite Time Lyapunov Exponent (FTLE), and to the 
Okubo-Weiss (OW) and Hua-Klein (HK) criteria. 
We show that the OW and HK are a limiting case of the 
FTLE, and that  the FSLE is the most efficient method for
detecting the presence of cross-stream barriers. 
We illustrate our findings by considering two examples of 
geophysical interest: a kinematic meandering jet model, and 
Lagrangian tracers advected by stratospheric circulation. 
\end{abstract}

%%%%%%%%%%%%%%%%%%%%%%%%%%%%%%%%%%%%%%%%%%%%%%%%%%%%%%%%%%%%%%%
\section{Introduction}
\label{sec:1}
%%%%%%%%%%%%%%%%%%%%%%%%%%%%%%%%%%%%%%%%%%%%%%%%%%%%%%%%%%%%%%%

Transport processes play a crucial role in many natural phenomena.
Among the many examples, we just mention particle transport in
geophysical flows which is of great interest for atmospheric and
oceanic issues. 
The most natural framework for investigating such processes is the 
Lagrangian viewpoint, in which the tracer
trajectory ${\bf x}(t)$ is advected by 
a given Eulerian velocity field ${\bf u}({\bf x},t)$ according to 
the differential equation
\begin{equation}
{d {\bf x} \over d t} = {\bf u}({\bf x},t) 
\label{eq:1.1}
\end{equation}
Despite of the simple formal relation (\ref{eq:1.1}), the problem of
connecting the Eulerian properties of ${\bf u}({\bf x},t)$ to the
Lagrangian properties of the trajectories ${\bf x}(t)$, and {\it viceversa},
is a very difficult task. In the last $20-30$ years the
scenario has become even more complex by the recognition of
the ubiquity of Lagrangian chaos (chaotic advection).
Even very simple Eulerian fields can generate unpredictable 
Lagrangian trajectories which are practically indistinguishable from
those obtained in a complex, turbulent, flow 
\cite{H66,Licht,Ottino,lagran}.

In the following we will restrict our attention to the case of two-dimensional
incompressible velocity field ${\bf u}({\bf x},t)$, with ${\bf x}=(x,y)$. 
Incompressibility
is automatically satisfied by introducing the stream function 
$\psi=\psi({\bf x},t)$
and consequently defining the velocity 
field in terms of partial derivatives as
${\bf u}=(\psi_{y},-\psi_{x})$.
The evolution equation (\ref{eq:1.1}) then becomes
\begin{equation}
{d x \over dt} = \psi_{y} \, , 
\hspace{1cm}
{d y \over dt}  = - \psi_{x} \,.
\label{eq:1.2}
\end{equation}
Formally (\ref{eq:1.2}) is a Hamiltonian system with Hamiltonian
$\psi({\bf x},t)$. Chaotic trajectories ${\bf x}(t)$ typically appear as a
consequence of the time dependence of $\psi$ \cite{Licht}.

Many geophysical flows, when observed at sufficiently large scale,
are within this class \cite{Samel,Yang}. Moreover time dependence can be often
considered a perturbation over a given steady flow, i.e. (\ref{eq:1.2})
represents a quasi-integrable Hamiltonian system.
It is well known that in quasi-integrable Hamiltonian systems chaos can be
quite non uniform in the phase space, due to the presence of the
invariant KAM tori with a chaotic layer around them \cite{Ottino}.
The presence of these regular islands (also called coherent structures
in the context of geophysical flows) is of particular importance for
the dispersion process because they act as barriers to transport.
The sensitivity of different diagnostics of transport to the 
presence of barriers will be the main topic of our investigation.
In particular, we will consider the Finite Time Lyapunov Exponent 
(FTLE), the Okubo-Weiss (OW) and Hua-Klein criteria (HK), and the local 
Finite Scale Lyapunov Exponent (FSLE).

We discuss these different methods by considering two examples. 
First, we study transport properties in a kinematic
meandering jet model, formerly introduced for 
describing the Gulf stream \cite{Bower,Samel}. 
Second, we analyze a large number of stratospheric isoentropic 
trajectories, computed according to (\ref{eq:1.1}) from assimilated 
wind fields, describing Lagrangian motion around the polar vortex. 
In both situations our results show that the existence of barriers limiting
the dispersion across the stream is well described by the FSLE but 
it is completely missed by the OK and HK criteria, 
which can only depict the landscape of alternating unstable hyperbolic and 
stable elliptic points of the flow. The FTLE will be 
discussed in relation to the OW and HK exponents and to the FSLE. 

The remaining of this paper is organized as follows. In 
Section \ref{sec:2} we introduce and discuss the different 
diagnostics for characterizing dispersion.
Section \ref{sec:3} is devoted to the analysis of the meandering jet  
and of the polar vortex. 
Finally, in Section \ref{sec:4} we present some conclusions.

%%%%%%%%%%%%%%%%%%%%%%%%%%%%%%%%%%%%%%%%%%%%%%%%%%%%%%%%%%%%%%%
\section{Characterization of local transport properties} 
\label{sec:2}
%%%%%%%%%%%%%%%%%%%%%%%%%%%%%%%%%%%%%%%%%%%%%%%%%%%%%%%%%%%%%%%

In presence of Lagrangian chaos, two close trajectories 
typically separate exponentially in time \cite{lagran}. 
Thus the natural statistics we adopt for describing chaotic 
particle spreading is the relative dispersion statistics.
Relative separation between two particles 
${\bf R}(t)={\bf x}^{'}(t)-{\bf x}(t)$ evolves according to
the velocity difference
\begin{equation}
{d {\bf R} \over d t} = {\bf u}({\bf x}(t)+{\bf R}(t),t) 
- {\bf u}({\bf x}(t),t) \, .
\label{eq:2.1}
\end{equation}
As far as particle separation remains much smaller than  
the typical length scale $l_{E}$ of the velocity field,
we can linearize (\ref{eq:2.1}) around the trajectory ${\bf x}$
and, for a generic time dependent flow, we expect exponential
growth of the separation, i.e.
\begin{equation}
R(t)  \simeq R(0) e^{\lambda t}
\label{eq:2.2}
\end{equation}
where $\lambda$ is the Lagrangian Lyapunov Exponent (LE).

In the opposite limit, $R \gg l_{E}$, the two particles feel uncorrelated
velocity fields and one recovers the standard diffusive regime, i.e.
\begin{equation}
\langle R^{2}(t) \rangle \simeq 2 D t
\label{eq:2.3}
\end{equation}
where the average is taken over many particle pairs and
where $D$ is the diffusion coefficient.

It is important to remark that, in most realistic situations, 
both asymptotic regimes, i.e. very small $R(t)$ for (\ref{eq:2.2})
and very large $R(t)$ for (\ref{eq:2.3}) cannot be attained.  
From one side, particles separation can be not sufficiently small
to justify the linearization leading to (\ref{eq:2.2}). In the
opposite limit, large separations cannot be reached in presence
of boundaries at scales comparable with $l_{E}$. 
As a consequence, the asymptotic quantities as $\lambda$ and $D$ 
cannot be computed and a non-asymptotic characterization of
transport is needed \cite{ABCCV97}.

Let us discuss, now, some techniques one can use to characterize 
local dynamical properties of a system, in particular the relative 
dispersion rate as a function of the initial position. 

%%%%%%%%%%%%%%%%%%%%%%%%%%%%%%%%%%%%%%%%%%%%%
\subsection{The Finite Time Lyapunov Exponent}
\label{sec:2.1}

Let us start from the definition of the Lagrangian
Lyapunov Exponent:
\begin{equation}
\lambda = \lim_{t \to \infty} \lim_{R(0) \to 0} {1 \over t}
\ln {R(t) \over R(0)}  
\label{eq:2.4}
\end{equation}
In (\ref{eq:2.4}) one basically assumes that the linearization 
of the perturbation $R(t)$ on a generic reference trajectory holds 
for an infinite time. This is correct only if the perturbation can
be considered infinitesimal at any time. 
The characteristic time naturally associated to the LE is known as 
the predictability time $T_{\lambda}=\lambda^{-1}$,  
which is the characteristic time at which one can predict the position
of the tracer in the future.
The FTLE is obtained by avoiding the limit $t \to \infty$ in (\ref{eq:2.4}).
This gives the instantaneous growth rate over a finite interval $\tau$ as
\begin{equation}
\gamma_{\tau}(t) = \lim_{R(t) \to 0} {1 \over \tau}  \ln {R(t+\tau) \over R(t)} 
\label{eq:2.5}
\end{equation} 
which, at variance with $\lambda$, depends on the initial point ${\bf x}(t)$.
The FTLE is distributed around a mean value which is nothing  
but the LE, $<\gamma_{\tau}> = \lambda$,  
where the average is computed over a virtually infinite
number of $\tau$ intervals along the trajectory \cite{BCFV01}.

In principle, even at very small $R(t)$, one must wait a certain 
time interval, $T_w$, needed to drive the perturbation along 
the Lyapunov direction \cite{GSO87}. 
This is necessary if we want to measure intrinsic properties 
of the system.  
In presence of many degrees of freedom, the possibility that 
the waiting time $T_w$ is of the same order of the predictability time 
$T_{\lambda}$ cannot be excluded (see \cite{BJPV98} for a review on 
the predictability time in extended systems). 

Let us now discuss some practical shortcomings arising when we want to 
analyze realistic situations described by experimental or model data.  
First, as we already rescaled, in quasi-integrable Hamiltonian systems, 
different regions in the phase space can display different behaviors 
\cite{Licht}.
As a consequence one has a non trivial spatial distribution of Lyapunov
exponents: zero if the trajectory lies in a regular island, positive if it 
diffuses across the stochastic layer. 
In special cases, when for instance structures of the velocity field 
are ``localized'' persistent features, at least within time intervals 
considerably longer than the characteristic Lagrangian time, 
a more refined description in terms of finite-time 
Lyapunov exponent can be more appropriate.

In order to measure $\gamma_{\tau}(t)$ at a point ${\bf x}(t)$ one
can make use of the following procedure. Backward integration in
time for an interval $T_*$ bring the trajectory at the point ${\bf x}(t-T_*)$.
An infinitesimal perturbation $\delta {\bf x}(t-T_*)$ is switched on
and it is integrated forward to $\delta {\bf x}(t)$.
It $T_*$ is sufficiently long, i.e. $T_* \ge T_{w}$, the perturbation
$\delta {\bf x}(t)$ will be aligned along the Lyapunov eigenvector and
a further integration to $\delta {\bf x}(t+\tau)$ will give the FTLE
according to (\ref{eq:2.5}) \cite{GSO87}.
In general, $T_w$ is not known {\it a priori} and can vary 
 with the initial conditions. More serious problems arise from  
the limits of resolution in the knowledge of experimental or simulated 
Lagrangian data, which  
 can easily disrupt the linear approximation scenario. 

When $R(t)$  attains finite sizes, i.e. is of the order of the 
characteristic lengths of the system, the so-called Finite-Scale 
Lyapunov Exponent (FSLE) give an appropriate description of the 
intrinsic physical properties of dispersion at different scales of motion. 
We will discuss this point in Section~\ref{sec:2.3}.

%%%%%%%%%%%%%%%%%%%%%%%%%%%%%%%%%%%%%%%%%%%%%
\subsection{The Okubo-Weiss and Hua-Klein criteria}
\label{sec:2.2}

In two-dimensional turbulent flows, the stirring properties of initially 
small material lines 
are related to the combined effect of eddy and jet features 
of the velocity field.  
In cases when continuous velocity fields  
are given, a popular way used to characterize the 
local rate of separation of initially close 
trajectories is the Okubo-Weiss (OW) criterion \cite{Okubo70,Weiss91}, 
based on the computation of the eigenvalues of the velocity gradient 
tensor. If explicit time dependence cannot be neglected, Hua and Klein (HK) 
\cite{HK98} have proposed a generalization of the OW  
criterion, based on the computation of the eigenvalue of the 
acceleration gradient tensor, related to the distribution of the pressure 
field. 

Let us recall the two criteria in the case of 2D incompressible 
velocity field with Lagrangian evolution given by (\ref{eq:1.2}).
The evolution of an infinitesimal separation ${\bf R}(t)$ is
given in the tangent space as
\begin{equation}
{d {\bf R} \over d t} = \left(
\begin{array}{ll}
\psi_{xy} & \psi_{yy} \\
-\psi_{xx} & -\psi_{xy}
\end{array}
\right)
{\bf R} \equiv {\bf M} {\bf R} \, ,
\label{eq:2.7}
\end{equation}
where the Jacobian ${\bf M}$ has the property ${\bf M}^2=\lambda_{0} {\bf 1}$
with $\lambda_{0}=-det({\bf M})$.
At small times, the solution of (\ref{eq:2.7}) is
\begin{equation}
{\bf R}(t) = \left[ {\bf 1} + {\bf M} t + {1 \over 2} {\bf N} t^2 \right] 
{\bf R}(0) 
+ O(t^3) 
\label{eq:2.8}
\end{equation}
where ${\bf N}=\lambda_{0} {\bf 1} + d{\bf M}/dt$.
The Okubo-Weiss criterion consists in computing $\lambda_0$, i.e. the product 
of the eigenvalueseigenvalues of ${\bf M}$. 
We recall that the quantity $\lambda_0$ can be written in terms of the
square strain $\sigma^2$ and the square vorticity $\omega^2$ as
\begin{equation}
\lambda_0 = {1 \over 4} (\sigma^2 -\omega^2)
\label{eq:2.9}
\end{equation}

If $\lambda_0$ is positive, the two eigenvalues of ${\bf M}$  are real, 
the velocity field is locally hyperbolic and strain overcomes rotation. 
For negative $\lambda_0$, we have imaginary eigenvalues and
a predominance of rotation over strain.

Of course, the Okubo-Weiss criterion may not 
 be sufficient to determine the local strain-vorticity balance 
in a time dependent flow. 
In this respect, the Hua-Klein criterion, being based on the sign
of the largest eigenvalue of the ${\bf N}$ matrix 
$\lambda_{+}=\lambda_0+ \lambda_1$, with 
\begin{equation}
\lambda_1 = \sqrt{{d{\psi_{xy}} \over dt}^2-
 {d {\psi_{xx}} \over dt} {d {\psi_{yy}} \over dt}}
\label{eq:HK}
\end{equation}
gives a ``more Lagrangian'' description.
Both of them provide a picture of the distribution 
of stable elliptic points and unstable hyperbolic points in the flow.
Let us observe that in the case of stationary velocity field,
$d{\bf M}/dt=0$ and one has $\lambda_{+}=\lambda_{0}$.

Let us remark the relationship existing between the HK criterion 
and the FTLE. The ``instantaneous'' Lyapunov exponent 
$\gamma_{\tau}$ estimates the growth rate of a typical perturbation 
within a finite time interval $\tau$, after the perturbation has 
aligned along the most unstable direction, so that it is an 
intrinsic property shared by all the trajectories (except for a set 
of zero probability measure). The HK eigenvalue $\lambda_{+}$ estimates the 
local maximum strain rate, regardless any alignment time of 
the perturbation, so that, from the Lagrangian point of view, it may be 
biased by transient behaviors. 
In other words, measuring $\lambda_{+}$ corresponds to measuring  
$\gamma_{\tau}$ at very small $\tau$ (the integration time step)
starting with a perturbation always aligned along the locally most 
unstable direction. 
In practical situations it can happen that the HK eigenvalue and the 
FTLE give similar local descriptions but, from a theoretical point 
of view, starting with a perturbation along the local most unstable 
direction is as arbitrary as choosing any other direction: after  
a transient, the time evolution will drive the (infinitesimal) perturbation 
definitely along the  most unstable Lyapunov  
eigenvector.    
       
%%%%%%%%%%%%%%%%%%%%%%%%%%%%%%%%%%%%%%%%%%%%%
\subsection{The local Finite Scale Lyapunov Exponent}
\label{sec:2.3}

In most cases of interest, the linear regime during which the exponential 
growth of the inter-particle distance occurs can be not resolved.
Particle spreading is generally observed on large spatial scales, 
of the order of the characteristic lengths of the system, and 
appropriate non linear techniques must be employed to quantify relative 
dispersion rates (see \cite{BCCLV00} for a review about 
non asymptotic properties of transport and mixing). 

Let us consider a very small (infinitesimal) initial perturbation
$R(0)$ on a trajectory ${\bf x}$.
For a chaotic system, after the initial transient, 
$R(t)$ typically grows exponentially in time 
according to (\ref{eq:2.2}).
We wait until $R(t)$ reaches a certain threshold $\delta_i$, at a given
time $t_i$. Let ${\bf x}_i$ the position of the trajectory 
${\bf x}$ at time $t_i$.
At a later time $t_f$, $R(t)$ will reach a larger threshold 
$\delta_f = r \cdot \delta_i$ with an assigned $r>1$. 
We now define the $r-$amplification time of $R(t)$, relatively to 
the initial position ${\bf x}_i$, as $\tau(\delta_i,{\bf x}_i, r)=t_f-t_i$.
From this quantity, the local Finite Scale Lyapunov Exponent is defined as:
\begin{equation} 
\lambda_r(\delta_i,{\bf x}_i) = {1 \over \tau(\delta_i,{\bf x}_i,r)} \ln r
\label{eq:2.10}
\end{equation}
The exponent $\lambda_r(\delta_i,{\bf x}_i)$ is a measure of the 
local amplification rate of a perturbation of size $\delta_i$ on a 
trajectory passing by the point ${\bf x}_i$. 
In the case of periodic time dependence in the equation of motion, 
the local FSLE will depend also on the initial phase.

The above described prescription is necessary in order to characterize 
the perturbation growth as an intrinsic property of the system. 
Of course, in realistic situations, we have to consider the following 
problems: the flow is not simply periodic and thus $\lambda_r$ may depend
explicitly on time. Moreover, we cannot observe distances below a 
certain threshold because of finite resolution, and thus the
initial perturbation cannot be considered infinitesimal.
Another important remark is that in practical applications it may not 
be possible to have information on a uniform distribution of initial 
conditions. In particular, let us consider a 2-D time dependent velocity 
field (e.g. the surface circulation of a sea or an isoentropic layer of 
the stratosphere) and the relative Lagrangian dispersion of trajectories. 
Usually one cannot set the initial distance between two particles to an 
arbitrarily small size, and wait until the relative 
position vector is aligned with the most unstable direction and starts 
expanding with a typical rate measured by the LE. 
Finite resolution imposes a lower 
limit to physically reasonable distance sizes, say $\delta$, and 
we can only hope to take into account all the possible 
realizations of the local dispersion rate by taking the average of    
$\lambda_r(\delta,{\bf x})$ ($r>1$) over a large number of directions around 
the initial point ${\bf x}$, i.e. on a sphere with radius $\delta$. 
Furthermore, if $\delta$ is not very small if compared to the characteristic 
lengths of the system, the linear regime of instability is already expired. 
The FSLE, by its nature, is a non linear indicator of trajectory instability 
so it can measure relative dispersion rates at finite scales of motion. 
How long can we follow two trajectories so that their FSLE is still 
meaningful as a local diagnostics ? 
The answer depends essentially on how much rapidly the velocity field varies 
in time relatively to the Lagrangian characteristic time. For instance, 
if $T_E$ is the time scale within which the Eulerian structures, e.g. 
current systems, change their geometrical and physical aspects, local FSLE's 
are meaningful only if they are observed on times 
$\ll T_E$, i.e. in an almost {\it frozen field} approximation.

Let us now briefly discuss the main objective of the present paper, i.e. the
detection of barriers in particle transport.
We seed the flow with an uniform distribution of Lagrangian tracers
and compute the FSLE for any trajectory according to the prescription
given above. 

If $\ell_{E}$ represents some characteristic length in the flow, e.g.
the typical size of the eddies around the Gulf Stream in the North Atlantic
Ocean or around the polar jet current in the stratosphere, then 
particles trapped inside vortices or traveling down jet streams may
never separates beyond the scale $\ell_{E}$, giving $\lambda_{r}(\delta)=0$
for $\delta \le \ell_{E}$. On the contrary, particles located in the 
chaotic layer will give a positive FSLE. Thus at an appropriate value of the
threshold $\delta$, the map of FSLE (\ref{eq:2.10}) can be used as 
an efficient indicator of the presence of barriers in transport.

We will see how a simple periodically perturbed meandering jet 
can be a significant test to show that a more appropriate diagnostics 
is required when macroscopic barriers are under investigation. 
An interesting result related to geophysical data analysis  
concerns the barrier effects of the jet current of the stratospheric 
polar vortex (southern hemisphere). 
This technique is being used also to study local mixing properties 
in ocean systems.

%%%%%%%%%%%%%%%%%%%%%%%%%%%%%%%%%%%%%%%%%%%%%%%%%%%%%%%%%%%%%%%
\section{Results}
\label{sec:3}
%%%%%%%%%%%%%%%%%%%%%%%%%%%%%%%%%%%%%%%%%%%%%%%%%%%%%%%%%%%%%%%

%%%%%%%%%%%%%%%%%%%%%%%%%%%%%%%%%%%%%%%%%%%%%
\subsection{Numerical experiments}

\begin{figure}
\centerline{\epsfig{figure=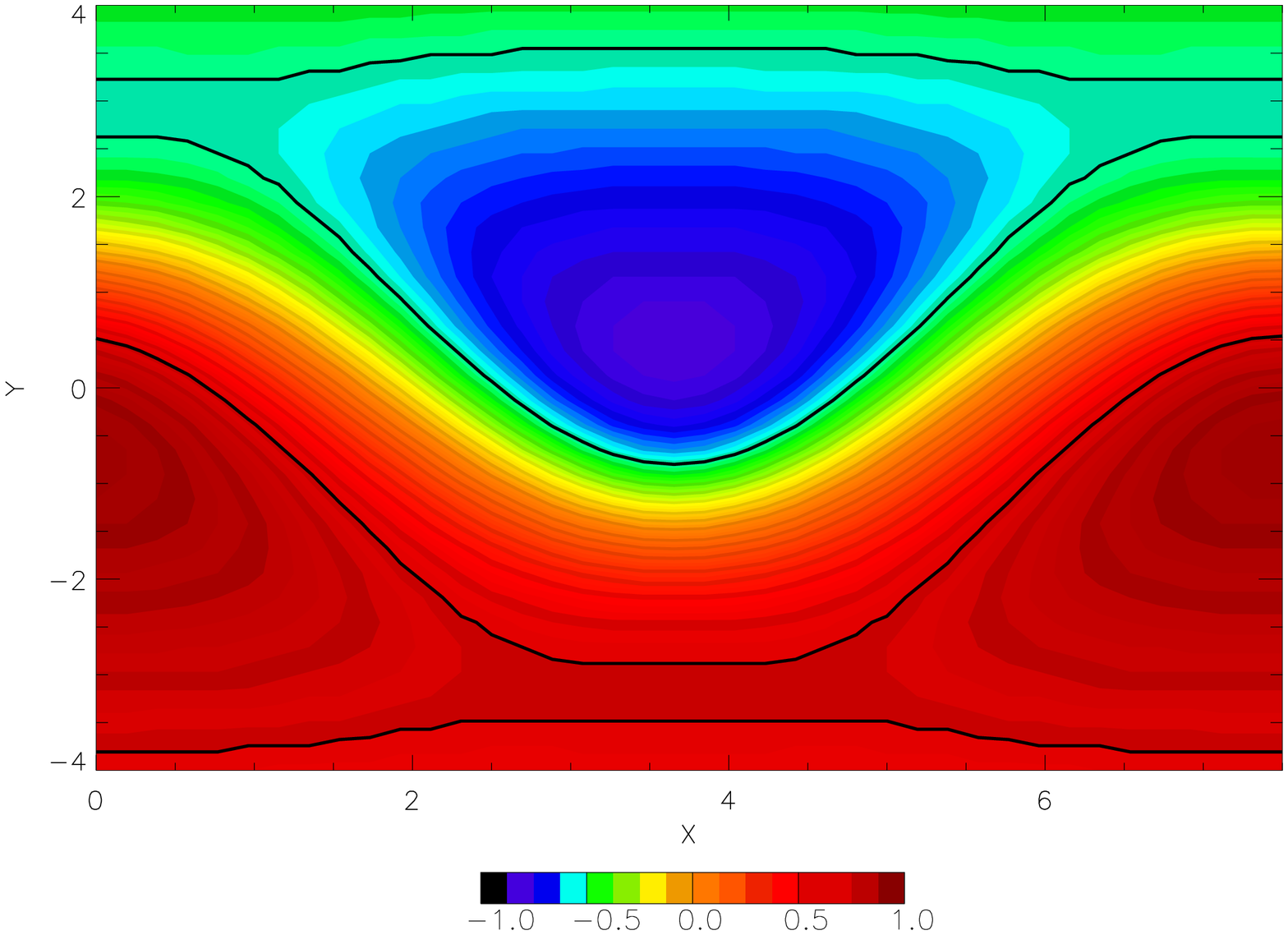,angle=90,width=12cm}}
\caption{Map of the stream function of the  
meandering jet model. The isolines drawn in black represent 
the borderlines of the current.   
The jet core becomes a permeable barrier 
to cross-stream motion depending on the parameter values 
$(\omega,\epsilon)$ of the time periodic perturbation.}
\label{fig:stream}
\end{figure}

We first discuss a simple, but not trivial, kinematic model in which  
a barrier to motion is known to exist for certain values of the 
parameters. The transition to the barrier-breakdown occurs by variation 
of some parameters.  We use this model to show what kind 
of information can be extracted from the different techniques 
previously discussed.
The system, formerly introduced as a model of transport across the 
Gulf Stream \cite{Bower,Samel}, consists a time periodic 
streamline pattern forming an oscillating meandering (westerly) current with 
recirculations along its boundaries:
\begin{equation}
\Psi= - tanh \left[ {y-B cos kx \over \sqrt{1 + k^2 B^2 sin^2 kx}} 
\right] + c y
\label{eq:psi}
\end{equation}
where $k$ is the spatial wave number of the structure, $c$ is the 
retrograde velocity in the ``far field'', $B$ is the amplitude of 
the meanders which varies periodically in time as
\begin{equation}
B=B_0 + \epsilon cos (\omega t + \phi)
\label{eq:B}
\end{equation}
The system can be fully mixing, i.e. any portion 
of the domain is definitely visited from any initial condition,
in a certain portion of the parameter space $(\epsilon, \omega)$
where in particular one has cross stream transport.
This system has two separatrices, with a spatial periodic structure, 
see Fig. \ref{fig:stream}, coinciding with the borderlines of the current. 
At very small $\epsilon$ the chaotic layer is restricted to a limited region 
around the separatrices, and no cross stream mixing occurs. In order to 
have large-scale chaotic mixing, i.e. particles jumping across 
the jet from a northern recirculation to a southern one, and vice-versa, 
one needs the overlap of the resonances \cite{Chirikov79}, when $\epsilon$ 
and $\omega$ are larger than certain critical thresholds.  

In Figure~\ref{fig:HKGULF} we report the OW indicator $\lambda_0({\bf x})$ 
as function of the initial position. Incidentally, the HK indicator 
$\lambda_+({\bf x}$ shows no significant differences from $\lambda_0$
and it is not shown. 
It is important to observe that both the OW and HK criteria 
are not able to detect the existence or not of a dynamical barrier,
i.e. the two figures are practically indistinguishable.

\begin{figure}
\centerline{a)}
\centerline{\epsfig{figure=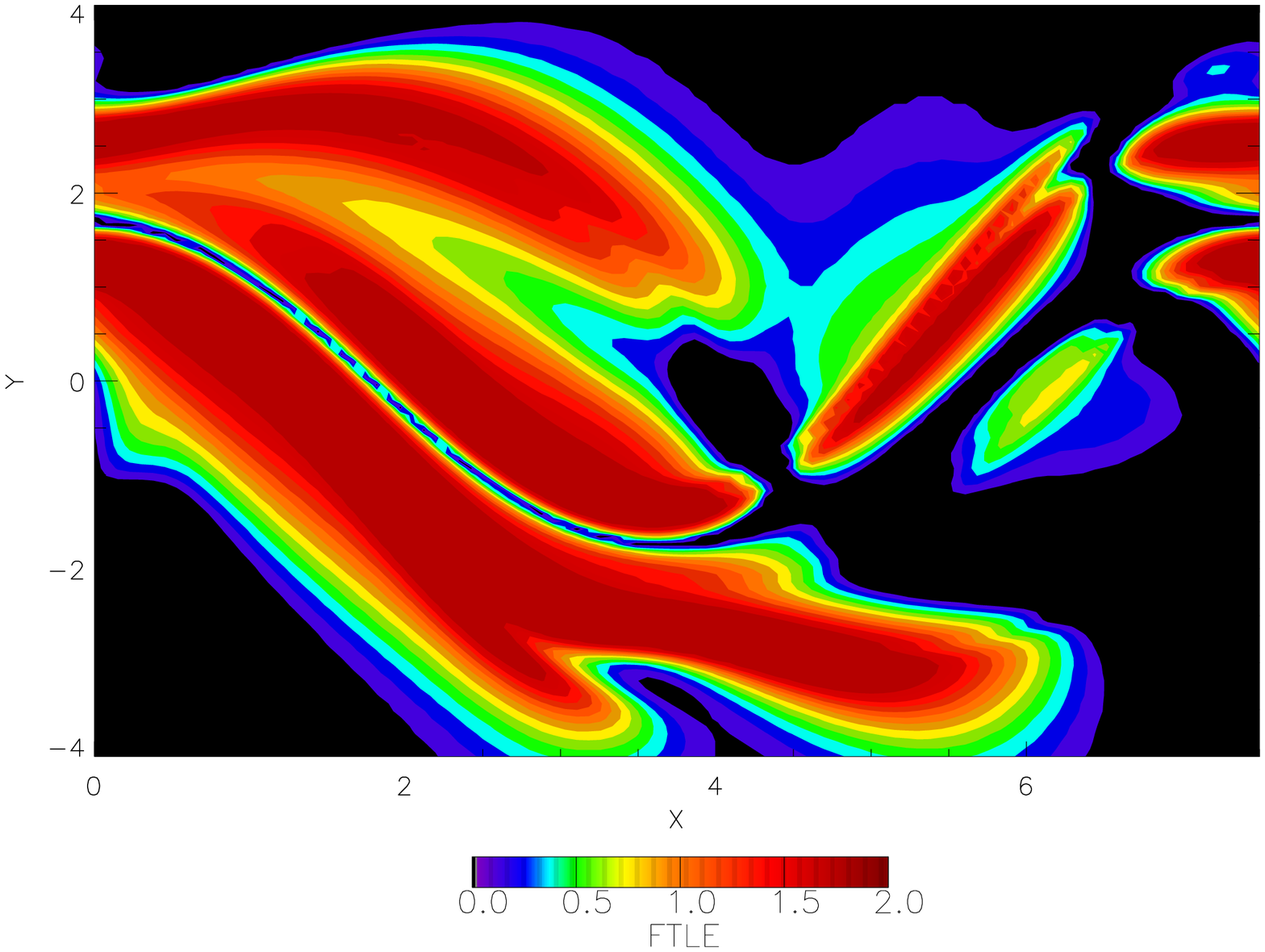,angle=90,width=12cm}}
\newpage
\centerline{b)}
\centerline{\epsfig{figure=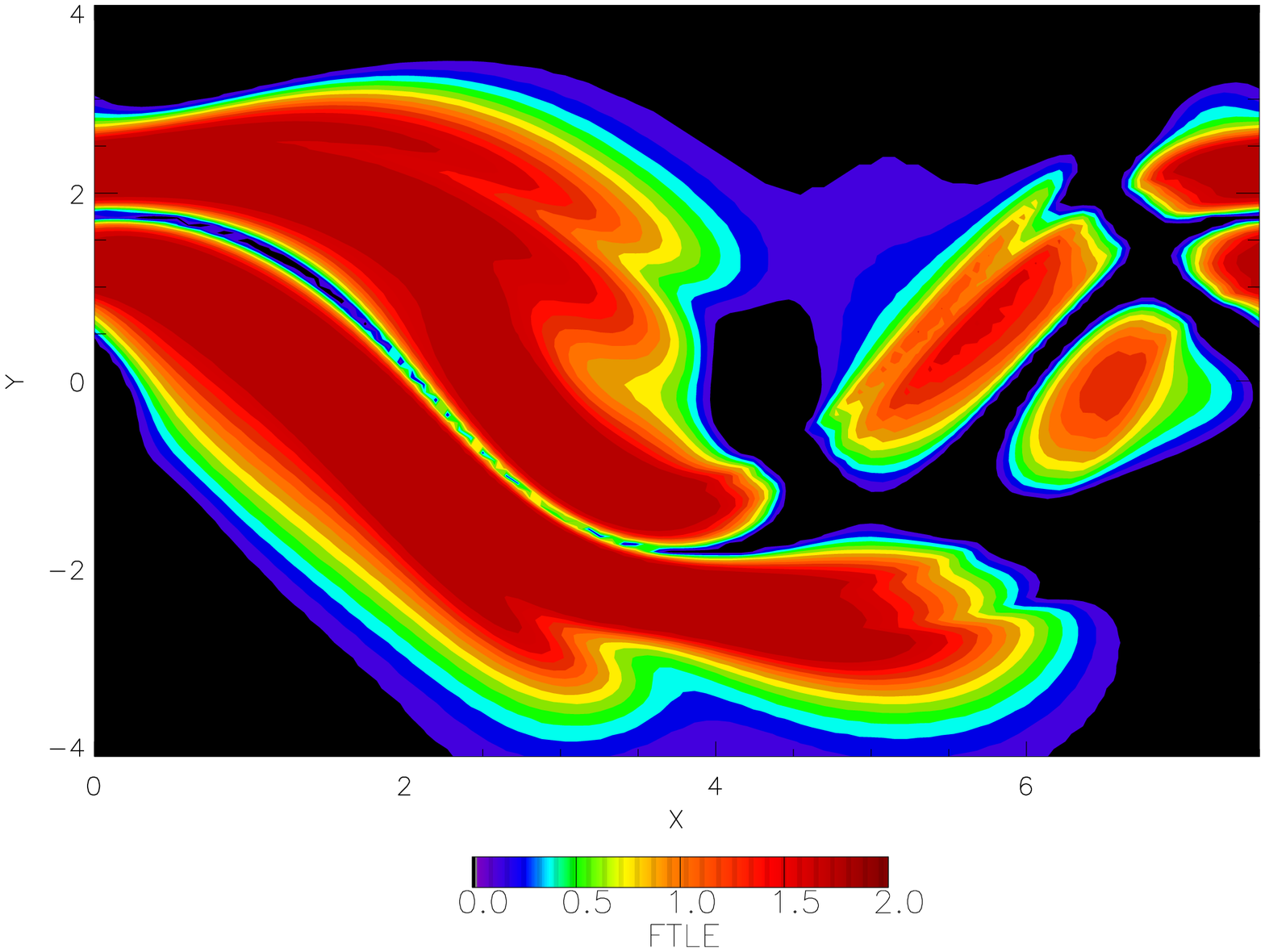,angle=90,width=12cm}}
\caption{Finite time Lyapunov exponent $\gamma_{\tau}({\bf x})$
for the meandering jet system at the two parameter points
$P_1$ (a) and $P_2$ (b). 
The number of particle pairs is $10000$ with initial separation
$\delta/L = 1.9 \times 10^{-3}$ uniformally distributed on the periodic
domain with spatial length $L=2 \pi/k$ (with $k=0.84$).
The time delay is $\tau=\pi/\omega_{0}$.}
\label{fig:FTGULF}
\end{figure}

The finite time Lyapunov exponent $\gamma_{\tau}({\bf x})$
is shown in Figure~\ref{fig:FTGULF}. The computation is
done according to (\ref{eq:2.5}) with initial separation
$R(t)=\delta$ for all the particle pairs, without the ``waiting 
procedure'' (i.e. $T_{w}=0$). This is because we want to
mimic a realistic situation of data analysis in which
arbitrarily small separations cannot be attained.

We note that, although the indicator is able to detect the
jet core (the low FTLE value filament inside the chaotic current),
the transition between the confined chaos regime and the large scale
mixing regime is not observed (compare Figure~\ref{fig:FTGULF}a and
Figure~\ref{fig:FTGULF}b).
The asymmetry of the FTLE map is due to the dependence of 
this indicator on the initial phase of the periodic flow. 

\begin{figure}
\centerline{a)}
\centerline{\epsfig{figure=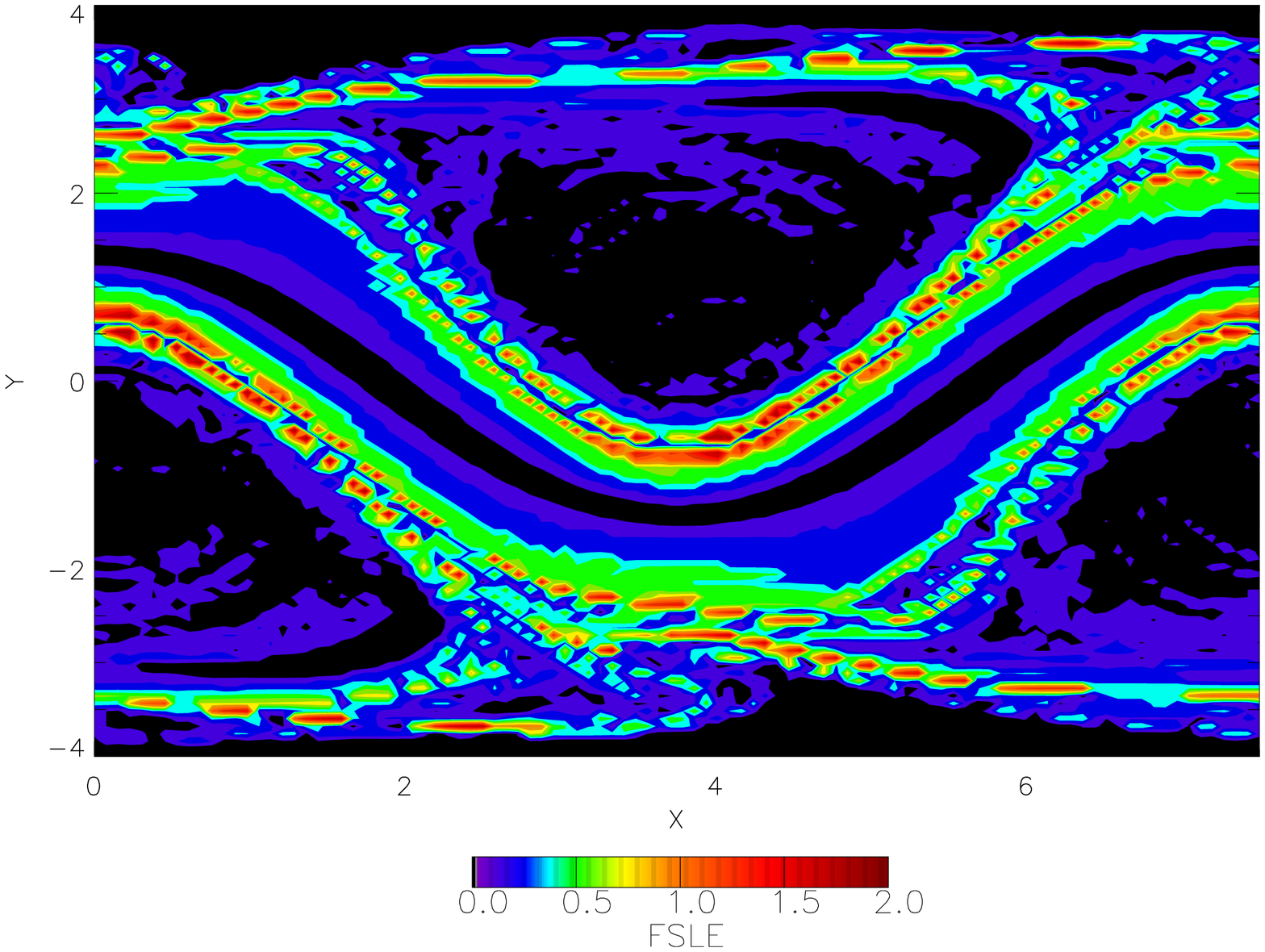,angle=90,width=12cm}}
\newpage
\centerline{b)}
\centerline{\epsfig{figure=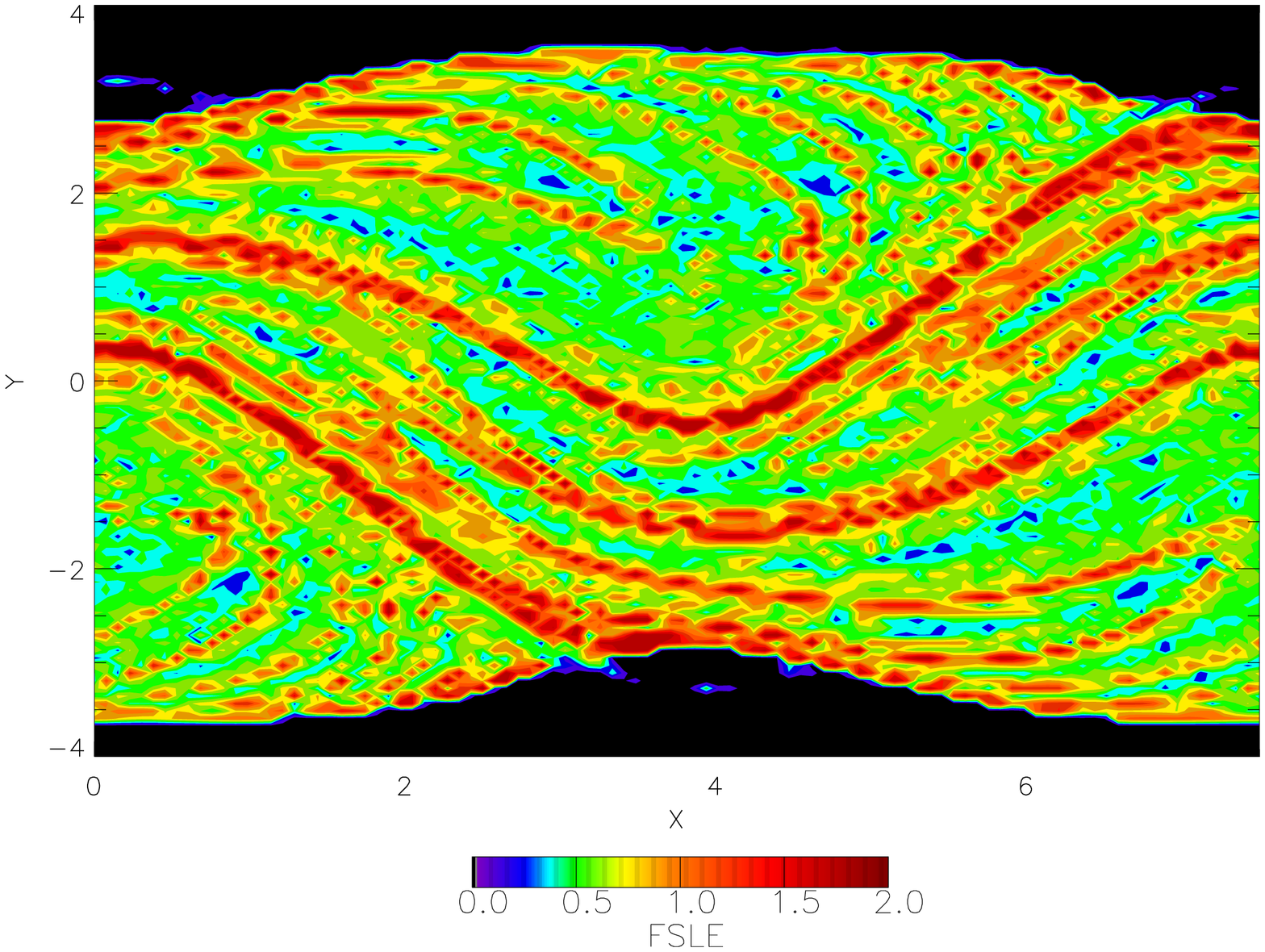,angle=90,width=12cm}}
\caption{The local FSLE $\lambda_r({\bf x})$ for the meandering jet system at 
the parameter space points $P_1$ (a) and $P_2$ (b).
The Lagrangian trajectories are the same of Figure~\ref{fig:FTGULF}.
The amplification factor is $r=100$. 
Only the particle pairs that reach a relative separation of
$r \cdot \delta \simeq 10^{-1} L$ give a 
positive signal in terms of $\lambda(r)$. 
} 
\label{fig:FSGULF}
\end{figure}

Let us consider the local FSLE as $\lambda_r({\bf x})$, computed
on the same trajectories and with the same initial separation 
of Figure~\ref{fig:FTGULF}.
Figure~\ref{fig:FSGULF}
contains the results of the $\lambda_r$ maps, before and after the 
overlap of the resonances, in what we can call 
the Melnikov \cite{Melnikov63} and the Chirikov \cite{Chirikov79} regimes, 
respectively. 
The amplification factor $r$ and the lower 
threshold $\delta$ are chosen such that 
the upper threshold $r \cdot \delta$ is of the order of the jet width. 
In one case, black 
regions of zero FSLE values, i.e. particle pairs which never reach the 
upper limiting separation, are located in the jetcore and in the 
centers of the recirculation: when chaos is still 
confined in the vicinity of the separatrices, no cross-stream 
transport is allowed. In the other case, after the separatrix breaking has 
occurred, no zero FSLE values are present, indicating that particles can 
spread apart over any distance from any initial condition. 

Let us conclude this section by remarking that a Lagrangian diagnostic
based on the FSLE shows is major skill in detecting dynamical 
barriers in the flow.

%%%%%%%%%%%%%%%%%%%%%%%%%%%%%%
\subsection{Geophysical data}

We now consider a geophysical example regarding Lagrangian motion on 
an isoentropic layer (i.e. at constant potential temperature) at 
a low stratospheric level, in presence of the winterly polar vortex 
\cite{schoeberl92}, characterized by a quasi-zonal robust jet stream.
 
\begin{figure}
\centerline{\epsfig{figure=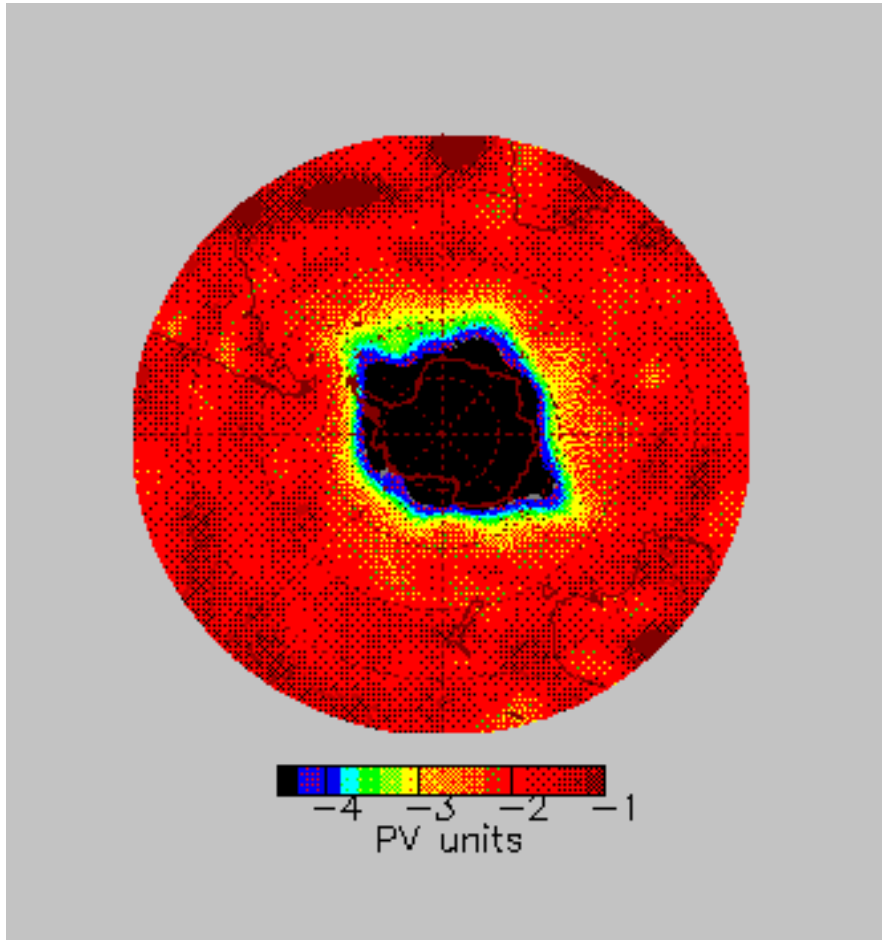,angle=0}}
\caption{Map of potential vorticity (PV) taken at day June 30th 1997, 
relative to the 475K layer, southern hemisphere.
}
\label{fig:PV}
\end{figure}

\begin{figure}
\centerline{\epsfig{figure=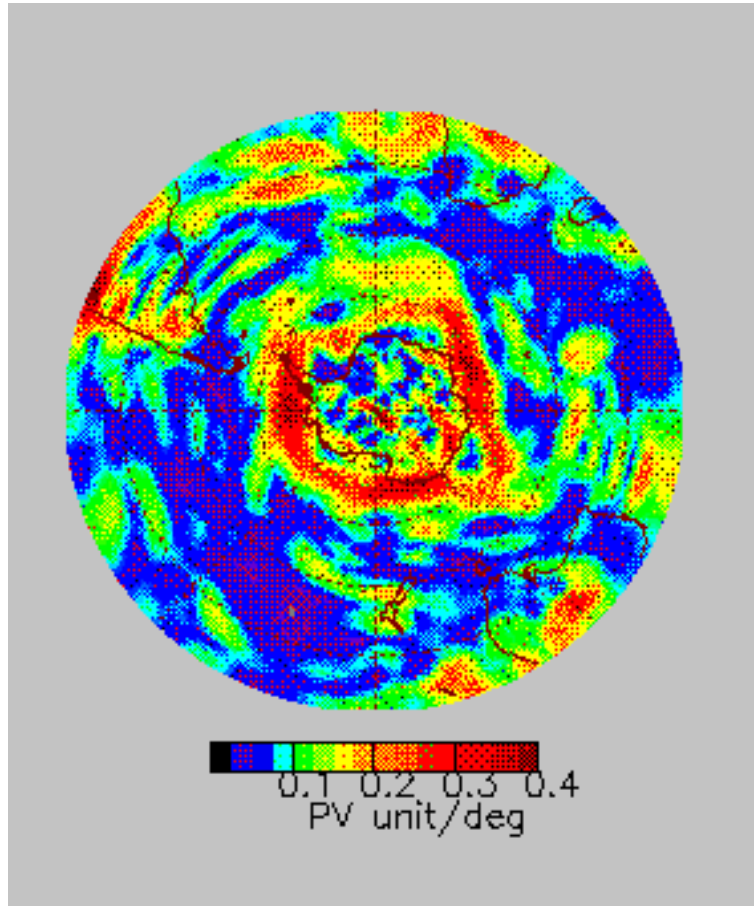,angle=0}}
\caption{Map of the magnitude of the gradient of potential vorticity
shown in Figure \ref{fig:PV}.}
\label{fig:GPV}
\end{figure}

The typical flow pattern is usually represented by means of 
stereographic maps of isoentropic Potential Vorticity (PV), 
see Figure~\ref{fig:PV}, and the modulus of its gradient shown in 
Figure~\ref{fig:GPV}.
In winter, the stratospheric PV can be considered as a quasi-conserved
quantity over a timescale of about 2-3 weeks. 
It is also widely accepted in literature that the outer border of
the polar vortex, usually identified by the maximum horizontal gradient 
of isoentropic PV \cite{Nash1996}, can act as a strong barrier to 
meridional cross stream transport.

The kinematics is remarkably similar to that of the previously 
discussed meandering jet model, if we imagine the latter as closed 
on itself in a circular geometry. 
The Lagrangian data set we have used for the analysis account of 
about $10^{4}$ trajectory pairs, initially uniformly distributed over the
southern hemisphere on the 475K isoentropic surface (lower stratosphere).
Trajectories are calculated by means of the University of L'Aquila 
Trajectory Model \cite{Redaelli1997,Dragani2000} using analyzed wind, 
pressure, and temperature fields from the U.K. Meteorological Office 
(UKMO) \cite{SN94} provided by the British Atmospheric Data Centre (BADC). 
Latitude coverage goes from poles to about tropics. 
The trajectories run from June 30th, 1997 up to a maximum observation 
time of 20 days. The initial distance between pair particles is 
$R_{i} \simeq 10 \, Km$. 

Being the trajectory evolution simulated with wind fields relative to 
the southern hemisphere winterly season, we are observing a situation of 
stable polar vortex regime \cite{SH91}. 

The local properties of the particle relative dispersion 
are obtained by computing the Okubo-Weiss 
eigenvalue $\lambda_{0}$ and the local FSLE $\lambda(r)$.
Spatial derivatives of the velocity fields used for the calculations are
estimated as finite differences over spatial grid steps of the order of 
$100 \, Km$.
 
\begin{figure}
\centerline{\epsfig{figure=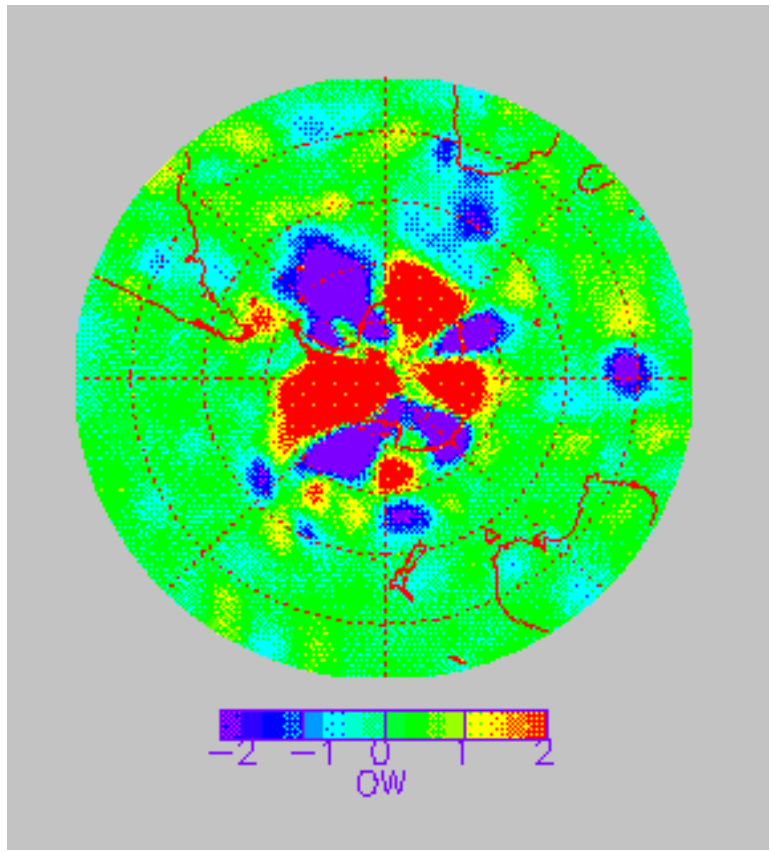,angle=0}}
\caption{Okubo-Weiss indicator $\lambda_0({\bf x})$ 
for the analyzed wind fields corresponding to Figure~\ref{fig:PV},
computed at the seventh day of the Lagrangian simulation, July 6th 1997. 
}
\label{fig:OWVORT}
\end{figure}

\begin{figure}
\centerline{\epsfig{figure=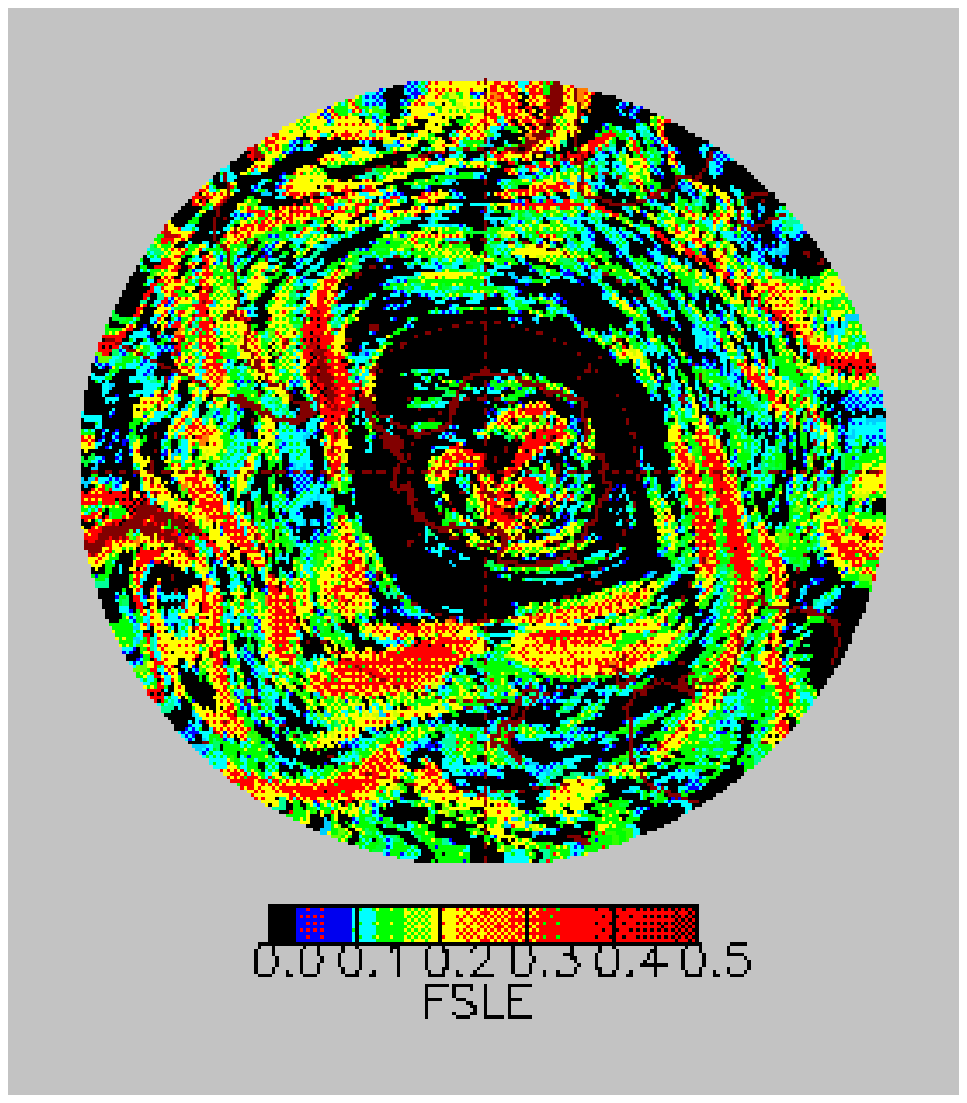,angle=0}}
\caption{The local FSLE $\lambda_r({\bf x})$ 
map for the isoentropic trajectory data set 
relative to the 475K layer. The initial distance 
between Lagrangian tracers is $\delta \simeq 10$ km and the 
amplification factor is $r=10$. 
}
\label{fig:FSVORT}
\end{figure}

The results are presented in the two-dimensional maps in
Figures \ref{fig:OWVORT} and \ref{fig:FSVORT}.
A further analysis (not shown) demonstrates that $\lambda_{0}$ does not 
change substantially in time and, as a consequence, it reproduces the essential 
features of the HK exponent $\lambda_{+}$. Neither of the exponents,
as in the previous example, is able to detect the presence of a 
barrier to transport. 

On the other hand, the FSLE map (see Figure~\ref{fig:FSVORT}) detects the
dynamical barrier as the region of vanishing FSLE values. The location 
of the barrier is in good agreement with the definition of polar vortex  
border based on geophysical considerations, e.g. the potential vorticity
gradient shown in  Figure \ref{fig:GPV} \cite{Nash1996}.

%%%%%%%%%%%%%%%%%%%%%%%%%%%%%%%%%%%%%%%%%%%%%%%%%%%%%%%%%%%%%%%
\section{Conclusions}
\label{sec:4}

We have discussed several techniques proposed for describing 
dispersion in two-dimensional flows. 
In particular our analysis has been focused on the capability
of these techniques to detect the presence of barriers to transport.
By means of two examples of geophysical relevance, we have shown
that Eulerian-based techniques, such as the Okubo-Weiss 
criterium and its generalization proposed by Hua and Klein, 
are not sensible to the presence of barriers. The Lagrangian 
finite-time Lyapunov exponent is, in principle, useful for describing 
space variations of the chaotic properties, e.g. in a quasi-integrable 
Hamiltonian system, but it is limited to small-scale properties of 
dispersion. 
A recent non-linear Lagrangian diagnostics, based on the Finite Scale
Lyapunov Exponent, is found to give the correct description
of the presence of large-scale barriers.  
As final remark, we notice that the OW criterion has been recently shown
to give poor information also in the case of fully developed 
turbulence \cite{rivera}: the probability distribution function 
of $\lambda_0$, $P(\lambda_0)$, for a typical 2D turbulent field is 
not sensitive to the presence of coherent structures, i.e. 
$P(\lambda_0)$ is the same as for a Gaussian field.

From a general point of view, it is not a surprise that 
purely Eulerian statistics, such as the OW quantity, are unable
to predict the behavior of Lagrangian tracers. The presence of
dynamical barriers is a fundamental information about the 
transport properties of the flow and thus can be considered
as a good discriminatory for the diagnostics.
It would be interesting to check the performance of the
proposed methods on other geophysical flows.

We thank the British Atmospheric Data Center for the UKMO data.
We are grateful to B. Joseph, D. Iudicone, B. Legras,  
R. Santoleri and G. Visconti for useful discussions.
This work is partially supported by MURST (contract N. 9908264583).

%%%%%%%%%%%%%%%%%%%%%%%%%%%%%%%%%%%%%%%%%%%%%%%%%%%%%%%%%%%%%%%

%%%%%%%%%%%%%%%%%%%%%%%%%%%%%%%%%%%%%%%%%%%%%%%%%%%%%%%%%%%%%%%
\newpage

%\centerline{FIGURE CAPTIONS}

\newpage
\begin{figure}
\centerline{\epsfig{figure=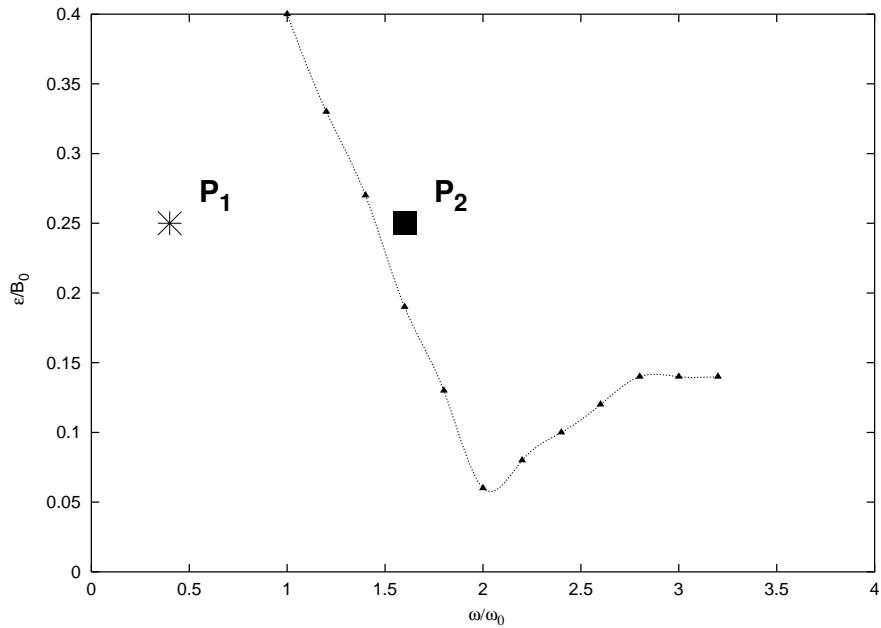,angle=-90,width=12cm}}
\caption{Critical curve in the parameter space $(\omega,\epsilon)$
separating between ``Melnikov regime'' (below the curve), in 
which chaos is confined close to the separatrices, and ``Chirikov 
regime'' (above the curve), in which large-scale chaotic mixing occurs. 
$P_1$, $(\omega,\epsilon)=(0.1,0.3)$, and $P_2$, $(\omega,\epsilon)=(0.4,0.3)$,
are the two points in the parameter space discussed 
in the Lagrangian simulations.
$\omega$ and $\epsilon$ are adimensionalized with respect to
$\omega_0=0.25$ (pulsation of recirculating orbits next to the 
separatrices) and $B_0=1.2$ (mean meander amplitude).}
\label{fig:overlap}
\end{figure}

\newpage
\begin{figure}
\centerline{a)}
\centerline{\epsfig{figure=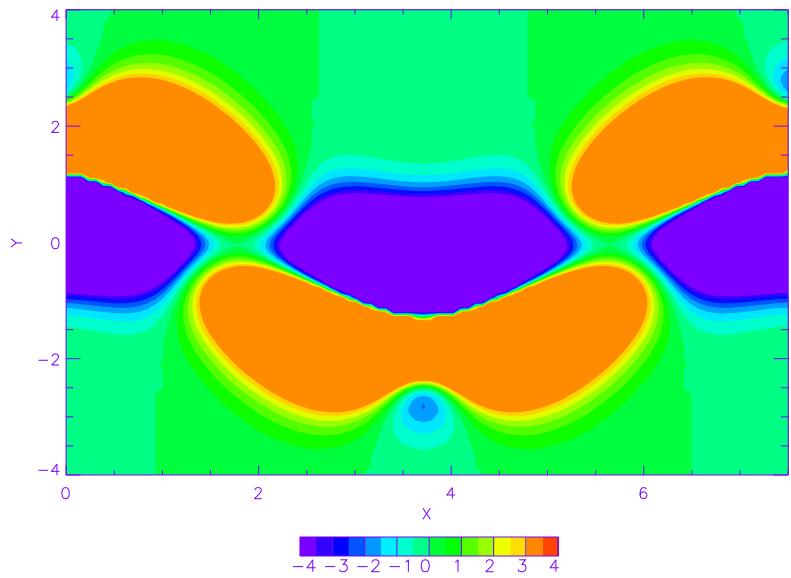,angle=90,width=12cm}}
\newpage
\centerline{b)}
\centerline{\epsfig{figure=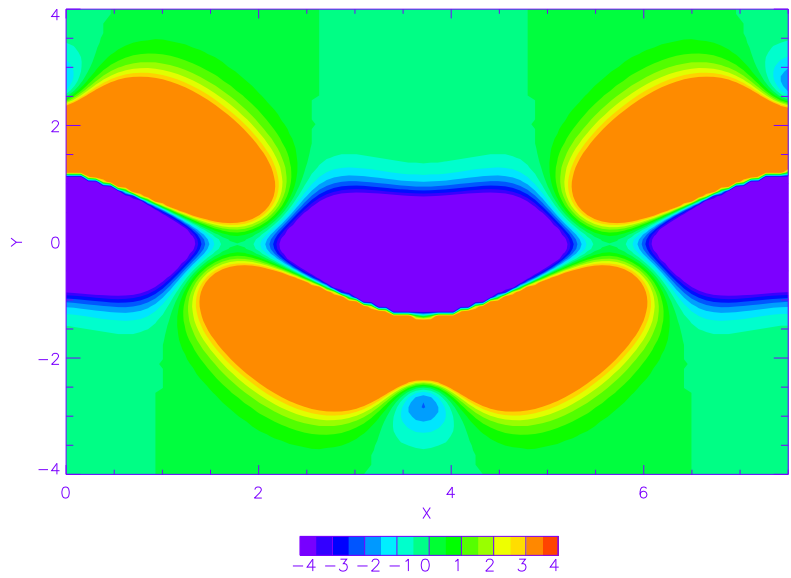,angle=90,width=12cm}}
\caption{Okubo-Weiss parameter $\lambda_0$ 
for the meandering jet system at 
the two parameter points $P_1$ (a) and $P_2$ (b) of 
Figure~\ref{fig:overlap}.}
\label{fig:HKGULF}
\end{figure}

\newpage

\newpage

\newpage

\newpage

\newpage

\newpage


\newpage
\begin{thebibliography}{99}

\bibitem{H66} M. H\'enon, 
Sur la Topologie des Lignes de courant dans in cas particulier,
C. R. Acad. Sci. Paris A 262 (1966) 312.

\bibitem{Licht} A.J. Lichtenberg and M.A. Lieberman,
Regular and Stochastic Motion (Springer, Berlin 1982).

\bibitem{Ottino} J.M. Ottino, 
The kinematics of mixing: stretching, chaos and transport
(Cambridge University Press, Cambridge 1989).

\bibitem{lagran} A. Crisanti, M. Falcioni, G. Paladin and A. Vulpiani,
Lagrangian Chaos: Transport, Mixing and Diffusion in Fluids,
Riv. Nuovo Cim. 14 (1991) 1.

\bibitem{Samel}
R.M. Samelson, 
Fluid exchange across a meandering jet,
J. Phys. Oceanogr. 22 (1992) 431.

\bibitem{Yang}
H. Yang,
Chaotic transport and mixing by ocean gyre circulation,
in Stochastic modeling in physical oceanography (ed. by R.J. Adler,
P. Muller and B.L. Rozovskiii) Birkh\"auser (1996) 439.

\bibitem{Bower}
A.S. Bower,
A simple kinematic mechanism for mixing fluid parcels across a 
meandering jet,
J. Phys. Oceanogr. 21 (1991) 173.

\bibitem{ABCCV97} V. Artale, G. Boffetta, A. Celani , M. Cencini and
A. Vulpiani,
Dispersion of passive tracers in closed basins:
beyond the diffusion coefficient, Phys. Fluids 9 (1997) 3162.

\bibitem{BJPV98}
T. Bohr, M. H. Jensen, G. Paladin and A. Vulpiani,
Dynamical systems approach to turbulence (Cambridge University Press,
Cambridge 1998).

\bibitem{HK98}
B.L. Hua and P. Klein,
An exact criterion for the stirring properties of nearly 
two-dimensional turbulence, 
Physica D 113 (1998) 98.

\bibitem{BCFV01} G. Boffetta, M. Cencini, M. Falcioni and
A. Vulpiani,
Predictability: a way to characterize complexity,
Phys. Report (submitted, 2001).

\bibitem{GSO87} I. Goldrisch, P.L. Sulem and S.A. Orszag,
Stability and Lyapunov stability of dynamical systems: a 
differential approach and a numerical method
Physica D 27 (1987) 311.

\bibitem{Okubo70} A. Okubo,
Horizontal dispersion of floatable particles in the vicinity
of velocity singularities such as covergences,
Deep-Sea Res. 17 (1970) 445.

\bibitem{Weiss91}
J. Weiss,
The dynamics of enstrophy transfer in two-dimensional
hydrodynamics,
Physica D 48 (1991) 273.

\bibitem{BCCLV00}
G. Boffetta, A. Celani , M. Cencini, G. Lacorata and A. Vulpiani,
Nonasymptotic properties of transport and mixing,
Chaos 10 (2000) 50.

\bibitem{Chirikov79}
B.V. Chirikov,
A universal instability of many-dimensional oscillator
systems,
Phys. Rep. 52 (1979) 263.

\bibitem{Melnikov63}
V.K. Melnikov,
On the stability of the center for time periodic perturbations,
Trans. Moscow Math. Soc. 12 (1963) 1.

\bibitem{rivera}
M. Rivera, X. L. Wu and C. Yeung,  
Universal Distribution of Centers and Saddles in Two-Dimensional Turbulence,
(preprint, arXiv:physics/0012051, 2001).

\bibitem{schoeberl92}
M.R. Schoeberl, L.R. Lait, P.A. Newman and J.E. Rosenfield, 
The structure of the polar vortex, 
J. Geophys. Res. 97 (1992) 7859.

\bibitem{SN94}
R. Swinbank and A. O'Neill, 
A stratosphere- troposphere data assimilation system, 
Mon. Weather Rev. 122 (1994) 686.

\bibitem{Redaelli1997}
G. Redaelli, 
Ph. D. Dissertation, University of L'Aquila, 1997. 

\bibitem{Nash1996}
E.R. Nash, P.A. Newman, J.E. Rosenfield and M. R. Scheberl,
An objective determination of the polar vortex using Ertel's potential
vorticity, 
J. Geophys. Res. 101 (1996) 9471.

\bibitem{SH91}
M. Schoeberl and Hartmann D. L., The dynamics of the 
Stratospheric Polar Vortex and its relation to springtime 
ozone depletions, Science 251 (1991) 45.

\bibitem{Dragani2000}
R. Dragani, G. Redaelli, G. Visconti, A. Mariotti, V. Rudakov, 
A. R. MacKenzie and L. Stefanutti, 
High resolution stratospheric tracer fields reconstructed with 
lagrangian techniques: a comparative analysis of predictive skill, 
J. of Atmos. Sci. (in press, 2001).

\end{thebibliography}
\end{document}